\documentclass[preprint,showpacs,showkeys,preprintnumbers,amsmath,amssymb,PRL,endfloats,floatfix]{revtex4}
\usepackage{graphicx}% Include figure files
\usepackage{dcolumn}% Align table columns on decimal point
\usepackage{bm}% bold math

\begin{document}

\preprint{arXiv}

\title{Electrically detected magnetic resonance using radio-frequency reflectometry}% Force line breaks with \\

\author{H.~Huebl}
\email[corresponding author
]{huebl@wmi.badw.de}\altaffiliation[present address:
]{Walther-Meissner-Institut, Bayerische Akademie der Wissenschaften,
Garching, Germany}\affiliation{Australian Research Council Centre of
Excellence for Quantum Computer Technology, School of Physics, The
University of New South Wales, Sydney, Australia}

\author{R.~P.~Starrett}
\affiliation{Australian Research Council Centre of Excellence for
Quantum Computer Technology, School of Physics, The University of
New South Wales, Sydney, Australia}

\author{D.~R.~McCamey}\altaffiliation[present address:
]{Department of Physics, University of Utah, 115 S 1400 E, Suite
201, Salt Lake City, Utah 84112-0830, USA} \affiliation{Australian
Research Council Centre of Excellence for Quantum Computer
Technology, School of Physics, The University of New South Wales,
Sydney, Australia}

\author{A.~J.~Ferguson}
\altaffiliation[present address:
]{Cavendish Laboratory, JJ Thomson Avenue, Cambridge CB3 0HE, United
Kingdom} \affiliation{Australian Research Council Centre of
Excellence for Quantum Computer Technology, School of Physics, The
University of New South Wales, Sydney, Australia}
\author{L.~H.~Willems~van~Beveren}
\affiliation{Australian Research Council Centre of Excellence for
Quantum Computer Technology, School of Physics, The University of
New South Wales, Sydney, Australia}

\date{\today}% It is always \today, today,

\begin{abstract}
The authors demonstrate readout of electrically detected magnetic
resonance at radio frequencies by means of an LCR tank circuit.
Applied to a silicon field-effect transistor at milli-kelvin
temperatures, this method shows a 25-fold increased signal-to-noise
ratio of the conduction band electron spin resonance and a higher
operational bandwidth of $>300$~kHz compared to the kHz bandwidth of
conventional readout techniques. This increase in temporal
resolution provides a method for future direct observations of  spin
dynamics in the electrical device characteristics.
\end{abstract}

\pacs{03.67.Lx, 72.20.-i, 73.40.Qv, 73.43.Fj}% PACS, the Physics and Astronomy

\keywords{silicon, phosphorus, Si, MOSFET ,tank circuit, spin, 2DEG,
electrically detected magnetic resonance, EDMR, ESR, Si:P,
quantum computing, rfEDMR, LCR, resonant circuit}%Use showkeys class option if keyword
\maketitle
% ========================= INTRODUCTION ============================================
Spin resonance is commonly used to spectroscopically identify spin
species in material systems, particular in semiconductors,
\cite{spaeth03} and it also allows spin dynamics to be studied using
pulsed excitation techniques.\cite{schweiger01} Although classical
spin resonance has the ultimate detection bandwidth, as the signal
is detected directly via the re-emission of microwave radiation, the
technique is usually limited to large spin ensembles.\cite{maier97}
In order to investigate few \cite{mccamey06} and ultimately a single
spin \cite{childress06, koppens06, wrachtrup93, koehler93} the spin
information is transferred to a secondary quantity, such as  charges
or photons, which can be detected sensitively. Such direct
electrical readout in electrically detected magnetic resonance
(EDMR) is typically limited to the low bandwidths in the kHz regime.
Therefore, coherent spin information can only be recovered from the
long term electronic evolution of the sample after the microwave
pulse.\cite{stegner06, huebl08} The development of radio frequency
(RF) detection schemes for single-electron transistor (SET)
electrometry \cite{schoelkopf98} has overcome the limitation of low
bandwidth. This has been achieved by embedding the sample in a
resonant LCR circuit, which matches the probed resistance
information to the impedance of the rf-circuitry of 50~$\Omega$,
allowing the measurement of the sample resistance at high
bandwidths.

In this letter we apply the LCR tank circuit developed for radio
frequency SETs to the detection of spin phenomena (in the following
referred to as rfEDMR). We measure the RF reflectance of the LCR
circuit as well as the low-frequency resistance of a
metal-oxide-semiconductor-field-effect-transistor (MOSFET), and
observe changes in both quantities under electron spin resonance.
Furthermore, the temporal dependence of rfEDMR is studied and
compared to conventional (low-frequency) EDMR demonstrating the high
bandwidth of rfEDMR. This could in principle allow the direct
investigation of coherently manipulated spin states, like Rabi
oscillations as opposed to the reconstruction discussed in detail in
Ref.~\cite{stegner06}.

Using a LCR resonant circuit for the detection of EDMR  requires
that the change in the device resistance, monitored at virtually DC
($<10~$kHz), can be detected at the resonance frequency $f_{RF}$ of
the LCR circuit, typically a few hundred MHz. While this seems valid
for high device mobilities, e.g. transport in the conduction band of
Si,\cite{graeff99} systems where the transport is hopping mediated
\cite{stutzmann00} may be an example where this method is not
applicable. For a silicon MOSFET, the spin-to-charge conversion
process originates from a difference in scattering amplitudes for
randomly formed spin pairs in singlet and triplet configuration in
the MOSFET channel, the relative contribution of which is changed
under spin resonance conditions, leading to a change in the total
conductance of the electron accumulation layer by $\Delta
\sigma/\sigma\approx 10^{-4}$.\cite{willemsvanbeveren08, ghosh92,
graeff99} Incorporating this device as the resistive element in a
LCR resonant circuit allows us to measure resistance changes via the
magnitude of the absorption at $f_{RF}$.\cite{schoelkopf98} This is
conveniently measured in reflection, where the reflection
coefficient $\Gamma$ is determined by the impedance mismatch between
the coaxial line and the LCR circuit. In EDMR spin resonance
manifests in a change $\Delta R$ of the total device resistance $R$
and for rfEDMR small changes in $\Delta R$, close to the matching
condition $|Z_{LCR}(f)|\approx50~\Omega$, will result in a change of
$\Gamma \propto \Delta R$. Additionally, calibration of the RF
reflectance can be obtained by measuring the DC device resistance
simultaneously and assuming that it is independent of the probe
frequency.\cite{schoelkopf98} The resistance can then be accessed
quantitatively using the RF circuit, retaining the benefit of the
increased bandwidth.
% ----------------------FIGURE -----------------------------------
\begin{figure}
\includegraphics[width=5.5cm]{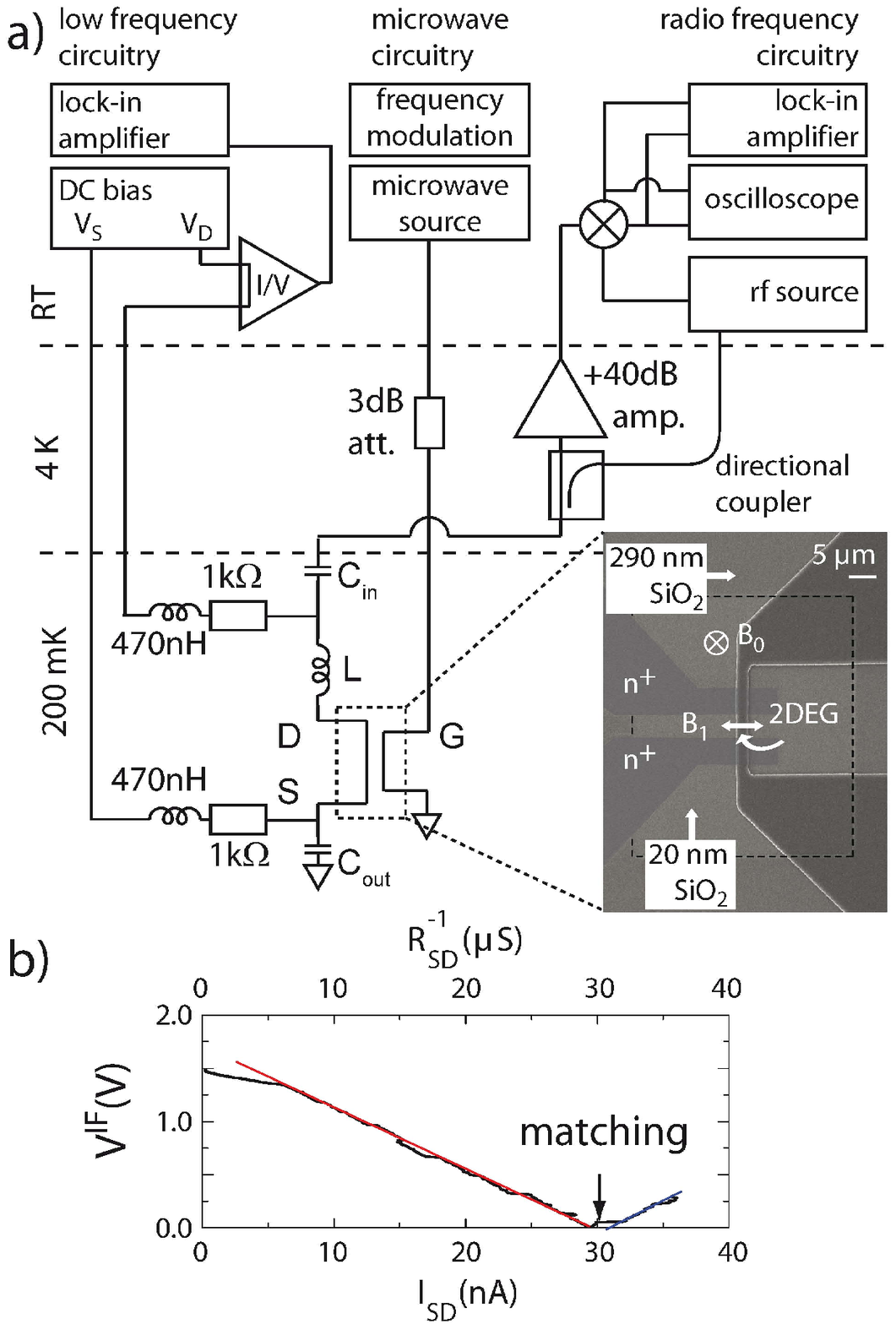}
\caption{(color online) Panel~(a) shows the EDMR detection setup
including the low frequency, the microwave, and the radio-frequency
circuitry. The inset shows a scanning electron micrograph of the
silicon MOSFET. Panel~(b) shows the calibration curve that
translates the source-drain conductance/current ($V_{SD}=$1~mV) to
the reflected RF signal amplitude after homodyning $V_{IF}$. At
30~nA a matching point is observed. The red and blue lines are
guides to the eye and show a slope of -57.9~mV/nA and 48~mV/nA,
respectively. \label{fig:rf-setup}}
\end{figure}
% ----------------------FIGURE -----------------------------------
%

The MOSFET (cf. Fig.~\ref{fig:rf-setup}~(a)) investigated is
fabricated on natural bulk silicon with a phosphorus doping of
$[{\rm{P}}]=8\times10^{16}$~cm$^{-3}$ and has a 20~nm thick,
high-quality SiO$_2$ gate oxide. Source and drain contacts are
formed by indiffused phosphorus regions with $[{\rm{P}}]=3\times
10^{19}$~cm$^{-3}$.\cite{willemsvanbeveren08} A shorted strip line
is used simultaneously as a gate electrode, to accumulate electrons,
and as a local antenna to provide the microwave magnetic field $B_1$
which excites the spin transitions. The MOSFET is turned on by
grounding the gate and applying a differential voltage across the
source and drain contacts $V_{SD}$, commonly offset from the top
gate potential $V_G$ and the source drain current $I_{SD}$ is
measured by a current amplifier (SIM918). For the experiments shown
here, the microwave frequency applied is $f_{MW}=37.9044~$GHz with a
source power of 10~dBm. In order to employ phase-sensitive detection
with a lock-in amplifier we modulate the frequency of the microwave
sinusoidally with variable rate $f_{mod}$ and a depth of 1.5~MHz
equivalent to 0.2~mT magnetic field
modulation.\cite{willemsvanbeveren08} The experiment is performed in
a dilution refrigerator at $\approx 200-300$~mK, which is inserted
into a superconducting magnet to provide a static magnetic field
$B_0$, oriented along the [001]-direction of the Si crystal and
perpendicular to the sample surface. Additionally, $B_0$ is
corrected for a constant offset, using the $^{31}$P hyperfine split
signature with a center of gravity $g$-factor of 1.9985 of a
reference spectrum.\cite{young97} Figure~\ref{fig:rf-setup}~(a)
shows the DC and RF part of the electrical setup  used for EDMR
detection in this experiment. The LCR circuit is realized by a
surface mount inductor ($L=330~$nH) on a printed circuit board
located close to the source contact of the MOSFET. The capacitance
($C\approx 500~$fF) is given by the parasitic capacitance from
source to drain and drain to the gate. The resistor $R$ is formed by
the source-to-drain resistance of the MOSFET. The resonance
frequency of this circuit under matching conditions is $385.13~$MHz
defining the RF frequency $f_{RF}$ used for rfEDMR. Additionally,
$C_{in}$ and $C_{out}$, each 100~pF, in combination with 1~k$\Omega$
NiCr resistors and 470~nH surface mount inductors form biasing tees,
decoupling the RF and DC signals. The RF reflectometry is realized
by sending an RF signal to a directional coupler (15~dB, at the 4~K
stage) and amplifying the reflected signal with a Quinstar cold
amplifier ($T_N=3.1~$K) located at the 4~K stage. After demodulating
the amplified reflected RF signal using a homodyne detection scheme,
involving quadrature demodulation (AD8348) at room temperature, the
resulting DC signal $V_{IF}\propto\Gamma$ is recorded by either a
lock-in amplifier or by a digital storage oscilloscope (DSO).

In order to calibrate the RF detection setup, we record the
source-drain current $I_{SD}$ as well as the demodulated RF signal
$V_{IF}$ for  $f_{RF}=385.13$~MHz as function of the top gate
voltage $V_{G}$. Figure~\ref{fig:rf-setup}~(b) shows $V_{IF}$ as
function of $I_{SD}$ or $1/R_{SD}$ for a source-drain bias
$V_{SD}=1$~mV. In the $I_{SD}$ regime between 6~nA and 30~nA, we
observe a linear relation of -57.9~mV/nA, which we use to convert
$V_{IF}$ into $I_{SD}^{IF}$ and $\Delta I_{SD}^{IF}$. The presence
of a matching point in the response of the tank circuit is a
particular advantage of rfEDMR, because here $\Gamma \propto \Delta
R$ which suppresses the finite $I_{SD}$ intrinsically.  In contrast,
conventional EDMR generally suffers from the challenge of measuring
a small resistance or current change on a relatively large current
offset.
%
% ----------------------FIGURE -----------------------------------
\begin{figure}
\includegraphics[width=5.0cm]{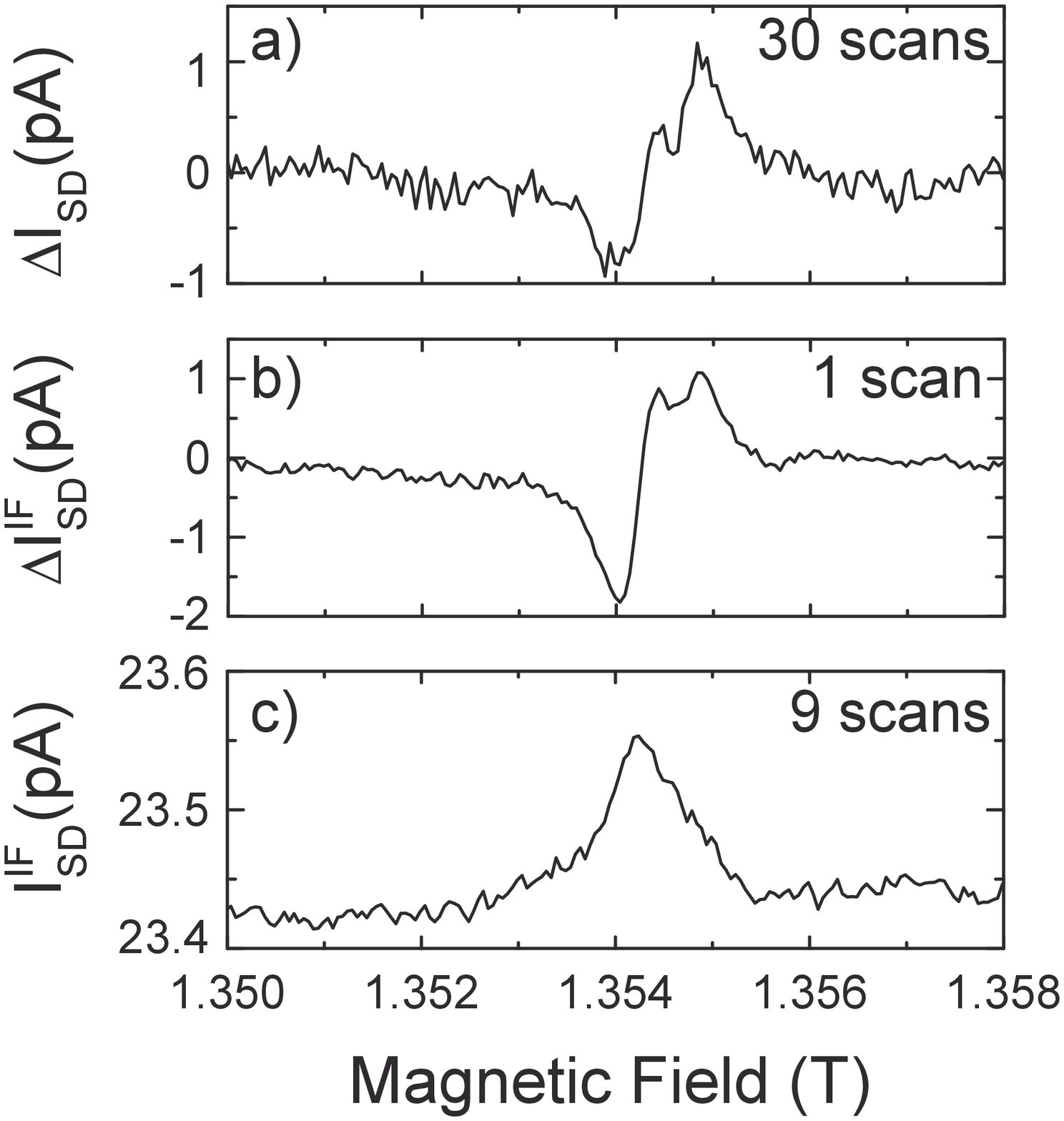}
\caption{EDMR using different detection schemes. Panel (a) shows the
EDMR signal obtained by monitoring the lock-in amplified source
drain current. Panel (b) shows the reflected RF signal from the LCR
circuit using homodyne detection, lock-in amplification, and
conversion to a current scale using Fig.~\ref{fig:rf-setup}~(b) with
a 25-fold increased signal-to-noise. Panel (c) shows the absolute
(converted) current change, monitored directly by measuring $V_{IF}$
with a storage oscilloscope. \label{fig:spectra}}
\end{figure}
% ----------------------FIGURE -----------------------------------
%

Figure~\ref{fig:spectra} compares the different EDMR detection modes
for a gate voltage $V_g=0.64$~V, a source-drain bias $V_{SD}=1$~mV
resulting in $I_{SD}=23.4$~nA, close to matching of the LCR circuit.
Panel~(a) shows the lock-in detected source drain current $\Delta
I_{SD}$ of a conventional EDMR spectrum acquired over a total of 30
magnetic field scans. Two resonant lines are observed, one at
$B_0=1.35429$~T corresponding to $g=1.9997$ in good agreement with
the conduction band electron signal of the accumulation layer
\cite{young97} and a second one at $B_0=1.35474~$T ($g=1.999$)
possibly originating from exchange coupled P donors.\cite{cullis75}
Interestingly, the two hyperfine-split lines expected for isolated
phosphorus donors in silicon are not present in this spectrum. Their
absence appears to be related to the bias conditions of the MOSFET,
as they are present for higher source-drain bias (50~mV) and lower
$V_G$ (0.3~V - not shown). This could be due to the trapping of
electrons in a diamagnetic P$^-$ state which is ESR inactive. This
state is approx 2-3~meV below the conduction band and therefore, the
thermal excitation at low temperatures $\propto 200-300~$mK and the
resulting slow escape rate could explain the absence of the
lines,\cite{morley08} whereas for high bias conditions an increased
escape rate due to tunneling is expected. This effect, whilst
interesting, is beyond the scope of this publication focussing on
the high-bandwidth readout, and will be discussed elsewhere.

Figure~\ref{fig:spectra}~(b) shows the demodulated and lock-in
amplified reflected RF signal, converted into an equivalent current
using the calibration curve of Fig.~\ref{fig:rf-setup}~(b) measured
under the same conditions as in Fig.~\ref{fig:spectra}~(a). This
spectrum was obtained in a single magnetic field scan with an
incident RF power of -55~dBm or $\sim0.4$~mV at the input of the LCR
circuit, similar to the dc bias applied. Fig~\ref{fig:spectra}~(b)
shows a significantly better signal-to-noise ratio of $S/N|_1=15$
per field scan in contrast to the conventionally acquired data shown
in (a) with $S/N|_1=0.65$ normalized to one scan. The improved
signal-to-noise ratio could originate from two effects: (i) at
higher probe frequencies in the 1/f regime the noise level of the
MOSFET is reduced \cite{sarpreshkar93} and (ii) the cold amplifier
has a better noise performance than the room temperature current
amplifier. This high signal-to-noise ratio enables monitoring the
device resistance sensitively enough to observe spin-dependent
resistance changes by direct acquisition of $V_{IF}$ with DSO.
Figure~\ref{fig:spectra}~(c) shows $V_{IF}$ converted to
$I^{IF}_{SD}$ using Fig.~\ref{fig:rf-setup}~(b) averaged over 9
magnetic field scans. Each field point consists of the mean value of
$V_{IF}$, obtained for a 1024-points time trace, where each trace is
averaged for 128 times. The ability to directly acquire
$I^{IF}_{SD}$ in the time domain with a precision that allows to
resolve spin-dependent features is the prerequisite for
investigating dynamic effects, e.g. resistance changes during or
after microwave pulses. In general, the detection bandwidth is
limited by the quality factor of the LCR circuit, in this case
20~MHz. This would technically allow the direct monitoring of Rabi
oscillations, since a typical Rabi frequency in pulsed ESR/EDMR
spectrometers is 5-20~MHz \cite{stegner06} and should be 2.8~MHz in
the devices used in these experiments, based on $B_1=0.1$~mT
obtained from numerical modeling.\cite{willemsvanbeveren08}

%
% ----------------------FIGURE -----------------------------------
\begin{figure}
\includegraphics[width=5.5cm]{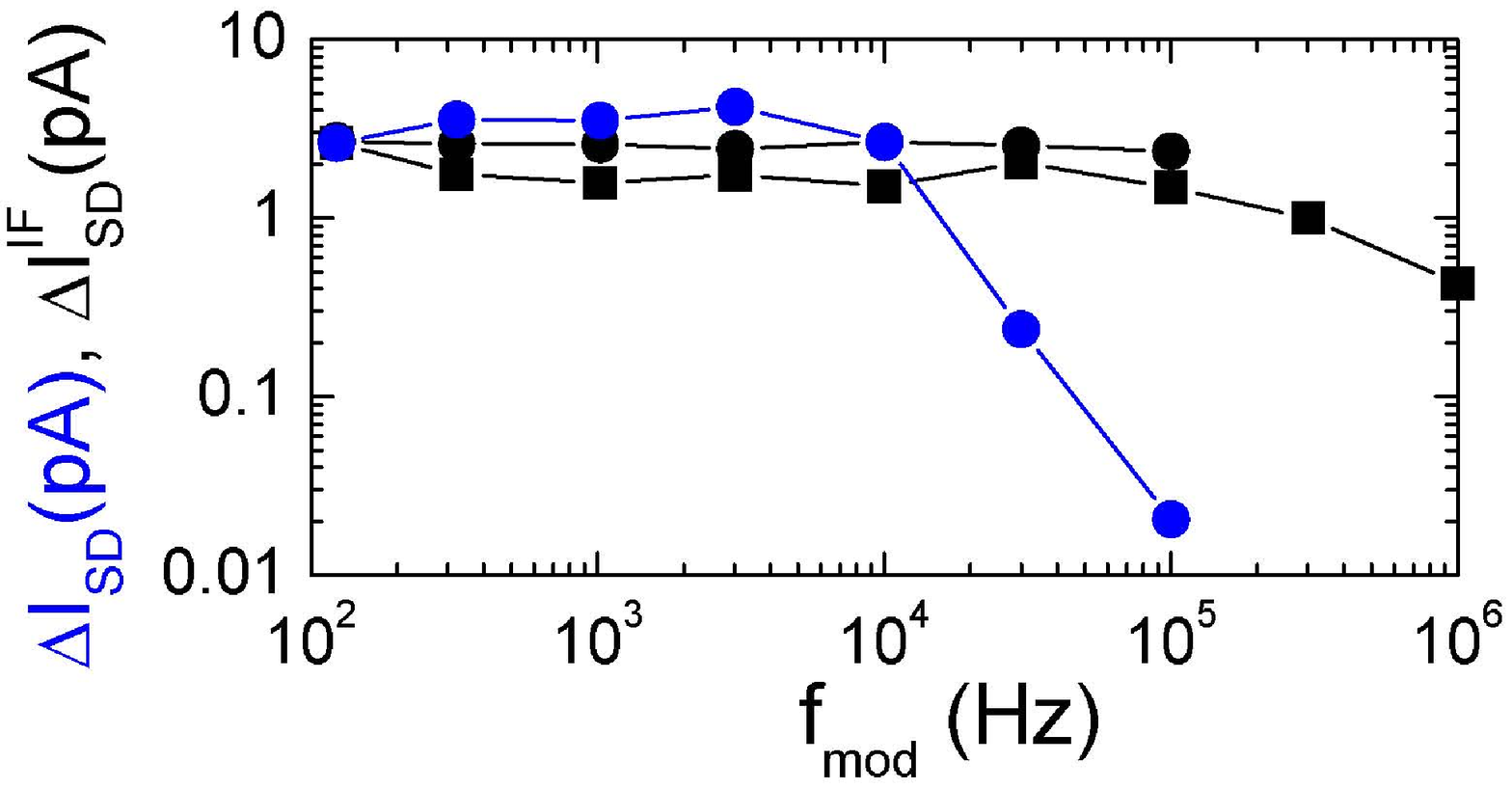}
\caption{Frequency modulation rate dependence of the signal
amplitudes for conventional EDMR and rfEDMR detection. The blue
circles represent the peak-to-peak amplitude of the conduction band
electron ESR signal using conventional EDMR, decreasing at $\approx
15$~kHz. The rfEDMR lock-in signal (black circles) matches the
conventional EDMR data up to $f_{mod}\approx 15$~kHz and shows no
decrease up to 100~kHz. Using a digital storage oscilloscope
$I^{IF}_{SD}$ was recorded up to $f_{mod}=1~$MHz and the lock-in
response determined numerically (black squares). Here, the signal
amplitude starts to drop at about 300~kHz, attributed to the voltage
preamplifiers used.\label{fig:freqdep}}
\end{figure}
%----------------------FIGURE -----------------------------------
%
To experimentally explore the bandwidth of rfEDMR, we investigate
the spin-resonant resistance change as function of the microwave
frequency modulation rate $f_{mod}$. Figure~\ref{fig:freqdep}
compares the conventionally measured (lock-in amplified)
peak-to-peak signal amplitudes obtained for the conduction band
electron signal at $B_0$=1.35429~T (blue squares) to the data
acquired via the reflected RF signal in the tank circuit, again
lock-in amplified (solid black circles). While the conventionally
detected signal shows a 3~dB compression at $f_{mod}\approx 15$~kHz,
as expected for a device resistance of $\approx$30~k$\Omega$ and
cable capacitance of $C_{cable}\approx 400$~pF resulting in a
$f_{RC}\approx~13~$kHz, the $\Delta I^{IF}_{SD}$ obtained via the
lock-in amplifier shows no frequency dependence up to 100~kHz, the
maximum frequency of the SR830 lock-in amplifier. To extend the
accessible frequency beyond this limitation, we directly digitized
$V_{IF}$ using a DSO and calculate the lock-in signal numerically
(solid black circles). For the digital data acquisition we kept the
number of recorded modulation periods constant. The response of the
LCR circuit measured with the SR830 and the 'numerical lock-in data'
agree quantitatively in the overlapping range, resulting in a
measured $f_{3dB}\approx300$~kHz. This limit is attributed to the
SR560 voltage preamplifiers used to post-amplify the output of the
quadrature demodulation board and could be overcome by higher
bandwidth preamplifiers.

In conclusion, we present a high bandwidth detection method to
observe EDMR using a LCR tank circuit. Measuring  rfEDMR for a
silicon MOSFET, quantitative agreement with conventionally detected
EDMR is demonstrated with a 25-fold increase in the signal-to-noise
ratio and a detection bandwidth $>300$~kHz. These two properties are
the key requirement to probe dynamic spin information in the
conductance directly, e.g. Rabi oscillations during ESR pulses.
Finally, this technique should be sensitive to capacitive changes in
the LCR circuit as well and might be applicable to capacitive
detected magnetic resonance studies.\cite{brandt00}

The authors thank D. Barber for technical support in the National
Magnet Laboratory at UNSW. This work is supported by the Australian
Research Council, the Australian Government, and by the U.S.
National Security Agency (NSA) and U.S. Army Research Office (ARO)
(under Contract No. W911NF-04-1-0290).

%\bibliography{HHmod2}% Produces the bibliography via BibTeX.

\begin{thebibliography}{20}
\expandafter\ifx\csname
natexlab\endcsname\relax\def\natexlab#1{#1}\fi
\expandafter\ifx\csname bibnamefont\endcsname\relax
  \def\bibnamefont#1{#1}\fi
\expandafter\ifx\csname bibfnamefont\endcsname\relax
  \def\bibfnamefont#1{#1}\fi
\expandafter\ifx\csname citenamefont\endcsname\relax
  \def\citenamefont#1{#1}\fi
\expandafter\ifx\csname url\endcsname\relax
  \def\url#1{\texttt{#1}}\fi
\expandafter\ifx\csname urlprefix\endcsname\relax\def\urlprefix{URL
}\fi \providecommand{\bibinfo}[2]{#2}
\providecommand{\eprint}[2][]{\url{#2}}

\bibitem[{\citenamefont{Spaeth and Overhof}(2003)}]{spaeth03}
\bibinfo{author}{\bibfnamefont{J.-M.} \bibnamefont{Spaeth}} \bibnamefont{and}
  \bibinfo{author}{\bibfnamefont{H.}~\bibnamefont{Overhof}},
  \emph{\bibinfo{title}{Point defects in Semiconductors and Insulators}}
  (\bibinfo{publisher}{Springer}, \bibinfo{address}{Berlin},
  \bibinfo{year}{2003}).

\bibitem[{\citenamefont{Schweiger and Jeschke}(2001)}]{schweiger01}
\bibinfo{author}{\bibfnamefont{A.}~\bibnamefont{Schweiger}} \bibnamefont{and}
  \bibinfo{author}{\bibfnamefont{G.}~\bibnamefont{Jeschke}},
  \emph{\bibinfo{title}{Principles of Pulse Electron Paramagnetic Resonance}}
  (\bibinfo{publisher}{Oxford University Press}, \bibinfo{address}{New York},
  \bibinfo{year}{2001}).

\bibitem[{\citenamefont{Maier}(1997)}]{maier97}
\bibinfo{author}{\bibfnamefont{D.~C.} \bibnamefont{Maier}},
  \bibinfo{journal}{Bruker Rep.} \textbf{\bibinfo{volume}{144}},
  \bibinfo{pages}{13} (\bibinfo{year}{1997}).

\bibitem[{\citenamefont{McCamey et~al.}(2006)\citenamefont{McCamey, Huebl,
  Brandt, Hutchison, McCallum, Clark, and Hamilton}}]{mccamey06}
\bibinfo{author}{\bibfnamefont{D.~R.} \bibnamefont{McCamey,}},  \bibnamefont{et
al.},
  \bibinfo{journal}{Appl. Phys. Lett.} \textbf{\bibinfo{volume}{89}},
  \bibinfo{pages}{182115} (\bibinfo{year}{2006}).

\bibitem[{\citenamefont{Childress et~al.}(2006)\citenamefont{Childress, Dutt,
  Taylor, Zibrov, Jelezko, Wrachtrup, Hemmer, and Lukin}}]{childress06}
\bibinfo{author}{\bibfnamefont{L.}~\bibnamefont{Childress,}},  \bibnamefont{et
al.},
  \bibinfo{journal}{Science} \textbf{\bibinfo{volume}{314}},
  \bibinfo{pages}{281} (\bibinfo{year}{2006}).

\bibitem[{\citenamefont{Koppens et~al.}(2006)\citenamefont{Koppens, Buizert,
  Tielrooij, Vink, Nowack, Meunier, Kouwenhoven, and Vandersypen}}]{koppens06}
\bibinfo{author}{\bibfnamefont{F.~H.~L.} \bibnamefont{Koppens}},  \bibnamefont{et
al.},
 \bibinfo{journal}{Nature}
  \textbf{\bibinfo{volume}{442}}, \bibinfo{pages}{776} (\bibinfo{year}{2006}).

\bibitem[{\citenamefont{Wrachtrup et~al.}(1993)\citenamefont{Wrachtrup, von
  Borczyskowski, Bernard, Orrit, and Brown}}]{wrachtrup93}
\bibinfo{author}{\bibfnamefont{J.}~\bibnamefont{Wrachtrup,}},  \bibnamefont{et
al.},
  \bibinfo{journal}{Nature} \textbf{\bibinfo{volume}{363}},
  \bibinfo{pages}{244} (\bibinfo{year}{1993}).

\bibitem[{\citenamefont{{K\"{o}hler} et~al.}(1993)\citenamefont{{K\"{o}hler},
  Disselhorst, Donckers, Groenen, Schmidt, and Moerner}}]{koehler93}
\bibinfo{author}{\bibfnamefont{J.}~\bibnamefont{{K\"{o}hler,}}},  \bibnamefont{et
al.},
  \bibinfo{journal}{Nature} \textbf{\bibinfo{volume}{363}},
  \bibinfo{pages}{242} (\bibinfo{year}{1993}).

\bibitem[{\citenamefont{Stegner et~al.}(2006)\citenamefont{Stegner, Boehme,
  Huebl, Stutzmann, Lips, and Brandt}}]{stegner06}
\bibinfo{author}{\bibfnamefont{A.~R.} \bibnamefont{Stegner}},  \bibnamefont{et
al.},
  \bibinfo{journal}{Nature Phys.} \textbf{\bibinfo{volume}{2}},
  \bibinfo{pages}{835} (\bibinfo{year}{2006}).

\bibitem[{\citenamefont{Huebl et~al.}(2008)\citenamefont{Huebl, Hoehne, Grolik,
  Stegner, Stutzmann, and Brandt}}]{huebl08}
\bibinfo{author}{\bibfnamefont{H.}~\bibnamefont{Huebl}},  \bibnamefont{et
al.},
\bibinfo{journal}{Phys. Rev. Lett.}
  \textbf{\bibinfo{volume}{100}}, \bibinfo{pages}{177602}
  (\bibinfo{year}{2008}).

\bibitem[{\citenamefont{Schoelkopf et~al.}(1998)\citenamefont{Schoelkopf,
  Wahlgren, Kozhevnikov, Delsing, and Prober}}]{schoelkopf98}
\bibinfo{author}{\bibfnamefont{R.~J.} \bibnamefont{Schoelkopf}},
  \bibnamefont{et al.},
  \bibinfo{journal}{Science} \textbf{\bibinfo{volume}{280}},
  \bibinfo{pages}{1238} (\bibinfo{year}{1998}).

\bibitem[{\citenamefont{Graeff et~al.}(1999)\citenamefont{Graeff, Brandt,
  Stutzmann, Holzmann, Abstreiter, and Sch\"{a}ffler}}]{graeff99}
\bibinfo{author}{\bibfnamefont{C.~F.~O.} \bibnamefont{Graeff}},
  \bibnamefont{et al.},
  \bibinfo{journal}{Phys. Rev. B} \textbf{\bibinfo{volume}{59}},
  \bibinfo{pages}{13242} (\bibinfo{year}{1999}).

\bibitem[{\citenamefont{Stutzmann et~al.}(2000)\citenamefont{Stutzmann, Brandt,
  and Bayerl}}]{stutzmann00}
\bibinfo{author}{\bibfnamefont{M.}~\bibnamefont{Stutzmann}},
  \bibinfo{author}{\bibfnamefont{M.~S.} \bibnamefont{Brandt}},
  \bibnamefont{and} \bibinfo{author}{\bibfnamefont{M.~W.}
  \bibnamefont{Bayerl}}, \bibinfo{journal}{Journal of Non-Crystalline Solids}
  \textbf{\bibinfo{volume}{266-269}}, \bibinfo{pages}{1}
  (\bibinfo{year}{2000}).

\bibitem[{\citenamefont{Willems van Beveren et~al.}(2008)\citenamefont{Willems van Beveren,
  Huebl, McCamey, Duty, Ferguson, Clark, and Brandt}}]{willemsvanbeveren08}
\bibinfo{author}{\bibfnamefont{L.~H.} \bibnamefont{Willems van Beveren}},  \bibnamefont{et
al.},
  \bibinfo{journal}{Appl. Phys. Lett.} \textbf{\bibinfo{volume}{93}},
  \bibinfo{pages}{072102} (\bibinfo{year}{2008}).

\bibitem[{\citenamefont{Ghosh and Silsbee}(1992)}]{ghosh92}
\bibinfo{author}{\bibfnamefont{R.~N.} \bibnamefont{Ghosh}} \bibnamefont{and}
  \bibinfo{author}{\bibfnamefont{R.~H.} \bibnamefont{Silsbee}},
  \bibinfo{journal}{Phys. Rev. B} \textbf{\bibinfo{volume}{46}},
  \bibinfo{pages}{12508} (\bibinfo{year}{1992}).

\bibitem[{\citenamefont{Young et~al.}(1997)\citenamefont{Young, Poindexter,
  Gerardi, Warren, and Keeble}}]{young97}
\bibinfo{author}{\bibfnamefont{C.~F.} \bibnamefont{Young}},  \bibnamefont{et al.}
 \bibinfo{journal}{Phys. Rev. B}
  \textbf{\bibinfo{volume}{55}}, \bibinfo{pages}{16245} (\bibinfo{year}{1997}).

\bibitem[{\citenamefont{Cullis and Marko}(1975)}]{cullis75}
\bibinfo{author}{\bibfnamefont{P.~R.} \bibnamefont{Cullis}} \bibnamefont{and}
  \bibinfo{author}{\bibfnamefont{J.~R.} \bibnamefont{Marko}},
  \bibinfo{journal}{Phys. Rev. B} \textbf{\bibinfo{volume}{11}},
  \bibinfo{pages}{4184} (\bibinfo{year}{1975}).

\bibitem[{\citenamefont{Morley et~al.}(2008)\citenamefont{Morley, McCamey,
  Seipel, Brunel, van Tol, and Boehme}}]{morley08}
\bibinfo{author}{\bibfnamefont{G.~W.} \bibnamefont{Morley}},  \bibnamefont{et al.}
  \bibinfo{journal}{Phys. Rev. Lett.} \textbf{\bibinfo{volume}{101}},
  \bibinfo{pages}{207602} (\bibinfo{year}{2008}).

\bibitem[{\citenamefont{Sarpeshkar et~al.}(1993)\citenamefont{Sarpeshkar,
  Delbr\"{u}ck, and Mead}}]{sarpreshkar93}
\bibinfo{author}{\bibfnamefont{R.}~\bibnamefont{Sarpeshkar}},
  \bibinfo{author}{\bibfnamefont{T.}~\bibnamefont{Delbr\"{u}ck}},
  \bibnamefont{and} \bibinfo{author}{\bibfnamefont{C.}~\bibnamefont{Mead}},
  \bibinfo{journal}{IEEE Circuits and Devices} \textbf{\bibinfo{volume}{9}},
  \bibinfo{pages}{23} (\bibinfo{year}{1993}).

\bibitem[{\citenamefont{Brandt et~al.}(2000)\citenamefont{Brandt, Neuberger,
  and Stutzmann}}]{brandt00}
\bibinfo{author}{\bibfnamefont{M.~S.} \bibnamefont{Brandt}},
  \bibinfo{author}{\bibfnamefont{R.~T.} \bibnamefont{Neuberger}},
  \bibnamefont{and}
  \bibinfo{author}{\bibfnamefont{M.}~\bibnamefont{Stutzmann}},
  \bibinfo{journal}{Appl. Phys. Lett.} \textbf{\bibinfo{volume}{76}},
  \bibinfo{pages}{1467} (\bibinfo{year}{2000}).

\end{thebibliography}

\clearpage
\end{document}